\documentclass[letterpaper, 10 pt, twocolumn]{article}  

\usepackage{graphics} 
\usepackage{epsfig} 
\usepackage{times} 
\usepackage{amsmath} 
\usepackage{amssymb}  
\usepackage{amsfonts}
\usepackage{graphicx}
\usepackage{epstopdf}
\usepackage{pdfpages}
\usepackage{upgreek}
\usepackage{mathptmx} 

\newcommand{\vectornorm}[1]{\left|\left|#1\right|\right|}

\usepackage{lipsum}

\newcommand\blfootnote[1]{%
  \begingroup
  \renewcommand\thefootnote{}\footnote{#1}%
  \addtocounter{footnote}{-1}%
  \endgroup
}

\title{Semi-global Output Feedback Stabilization of Non-Minimum Phase Nonlinear Systems}

\author{Almuatazbellah M. Boker and Hassan K. Khalil \blfootnote{This work was supported in part by NSF under grant number ECCS-1128467. A preliminary version of the material in this paper was presented at ACC 2013.}
\blfootnote{A. Boker is with the Department of Electrical Engineering, Marshall University, Huntington, WV 25755, bokera@marshall.edu.}%
\blfootnote{H. Khalil is with the Department of Electrical and Computer Engineering, Michigan State University, East Lansing, MI 48823, USA, khalil@msu.edu}}

\date{}
\begin{document}

\maketitle
\thispagestyle{empty}
\pagestyle{empty}

\begin{abstract}
We solve the problem of output feedback stabilization of a class of nonlinear systems, which may have unstable zero dynamics. We allow for any globally stabilizing full state feedback control scheme to be used as long as it satisfies a particular ISS condition. We show semi-global stability of the origin of the closed-loop system and also the recovery of the performance of an auxiliary system using a full-order observer. This observer is based on the use of an extended high-gain observer to provide estimates of the output and its derivatives plus a signal used by an extended Kalman filter to provide estimates of the remaining states. Finally, we provide a simulation example that illustrates the design procedure.
\end{abstract}

\section{Introduction}

There have been many results regarding the problem of output feedback stabilization of nonlinear systems. In recent years, more attention was directed towards the study of non-minimum phase nonlinear systems. One of the early papers to study stabilization of non-minimum phase nonlinear systems is \cite{isidori2000tool}. This paper proves semi-global stabilization for a general class of non-minimum phase nonlinear systems assuming the existence of a dynamic stabilizing controller for an auxiliary system. The same problem was considered in \cite{nazrulla2011robust}, where robust semi-global stabilization was achieved under similar assumptions but with an extended-high-gain-observer-based output feedback controller.

In \cite{karagiannis2005output}, \cite{marino2005class}, \cite{andrieu2008global} and \cite{ding2005semi} the problem of output feedback stabilization for nonlinear systems, which may have unstable zero dynamics, is solved. Paper \cite{karagiannis2005output} allows model uncertainty and is based on the observer introduced in \cite{karagiannis2003two}. Paper \cite{marino2005class} assumes the system will be minimum phase with respect to a new output, defined as a linear combination of the state variables. Paper \cite{andrieu2008global} assumes the knowledge of an observer and provides different approaches to design the control law. Papers \cite{karagiannis2005output}, \cite{marino2005class} and \cite{andrieu2008global} achieve global stabilization results under various stabilizability conditions on the internal dynamics. Paper \cite{ding2005semi} solves the stabilization problem for systems with linear zero dynamics. It uses the backstepping technique and an observer with linear error dynamics to achieve semi-global stabilization. Another result reported in \cite{hoseini2010adaptive} deals with a special case of the normal form where the internal dynamics are modelled as a chain of integrators. It uses an adaptive output feedback controller, based on Neural Networks and a linear state observer, to achieve ultimate boundedness of the states in the presence of model uncertainties.

In this work, we make use of the Extended Kalman Filter-Extended High Gain Observer (EKF-EHGO) approach, proposed in \cite{boker2013nonlinear}, to solve the problem of output feedback stabilization. In \cite{boker2013nonlinear}, we showed that we could use the extended high-gain observer to provide the derivatives of the measured output plus an extra derivative. This extra state is then used to provide a virtual output that can be used to make the internal dynamics observable. An extended Kalman filter can then use the virtual output to provide estimates of the remaining states; this is indeed possible thanks to the difference in time scale provided by the EHGO. The advantage of this observer is that it allows us to design the feedback control as if all the state variables were available. This is in contrast to previous high-gain observer results, which were mostly limited to partial state feedback and hence only to minimum phase systems.

In \cite{boker2013nonlinear}, we considered systems in the general normal form and proved local convergence for the EKF-EHGO observer. In the special case when the system is linear in the internal states, we proved a semi-global convergence result. The contribution of this paper is related to this special case.\footnote{In this paper, however, we allow the system to depend on the control in a more general manner.} We show that when the observer is used in feedback control we can achieve semi-global stabilization. We allow the use of any globally stabilizing state feedback control on the assumption that it satisfies an ISS condition with respect to the estimation error of the internal states. Furthermore, the class of systems considered includes non-minimum-phase systems. The proposed controller has the ability to recover the state feedback response of an auxiliary system that is based on the original system. This is beneficial as we can shape the response of the auxiliary system using the state feedback control as desired, with the knowledge that we will be able to recover it using the proposed output feedback.

The rest of the paper is organized as follows. Section \ref{s0} formulates the problem, presents the state feedback control and the extended Kalman filter, and describes the use of saturation to overcome peaking.   In Section \ref{s1}, we present the output feedback control and state two theorems that demonstrate the stability and trajectory recovery properties of the closed-loop system. Section \ref{ex} presents an example that illustrates the design procedure. Finally, Section \ref{s6} includes some concluding remarks.
\section{Preliminaries} \label{s0}
\subsection{Problem Formulation}
We consider a nonlinear system of the form
\begin{align}
&\dot{\eta}=A_1(\xi,u)\eta+\phi_0(\xi,u),\label{13}\\
&\dot{\xi}=A\xi+ B[C_1(\xi,u)\eta+a(\xi,u)],\label{23}\\
&y = C\xi,\label{33}
\end{align}
where $\eta \in R^{n-\rho},\xi \in R^{\rho}$, $y \in R$ is the measured output, and $u \in R$ is the control input. The $\rho \times \rho$ matrix $A$, the $\rho \times 1$ matrix $B$ and the $1 \times \rho$ matrix $C$ represent a chain of $\rho$ integrators and the functions $A_1, \phi_0, C_1$ and $a$ are known.

In the case that the nonlinear functions $A_1, \phi_0$ and $C_1$ are independent of the control $u$, \eqref{13}-\eqref{33} is a special case of the normal form \cite{isidori1997nonlinear}, where the system dynamics are linear in the internal state $\eta$. An example of systems of this type is the Translating Oscillator with a Rotating Actuator (TORA) system \cite{tora}, \cite{boker2013nonlinear}.

\newtheorem{assumption}{Assumption}
\begin{assumption}\label{ass1}
The functions $A_1$, $\phi_0$, $a$ and $C_1$ are sufficiently smooth. Furthermore, $\phi_0(0,0)=0$ and $a(0,0)=0$.
\end{assumption}

The goal is to stabilize the origin of the system \eqref{13}-\eqref{33} using only the measured output $y$.

\subsection{State Feedback}
We consider a state feedback control of the form
\begin{equation}
u=\gamma(\eta,\xi). \label{u3}
\end{equation}
Rewriting the full state vector as $\vartheta=[\eta^T,\; \xi^T]^T$,  the closed loop system under \eqref{u3} takes the form
\begin{equation}
\dot{\vartheta}=f(\vartheta,\gamma(\eta,\xi)).\label{closed}
\end{equation}
The function $\gamma$ is required to have the following properties.
\begin{assumption}\label{ass23}
\begin{enumerate}
	\item $\gamma$ is continuously differentiable with locally Lipschitz derivatives and $\gamma(0,0)=0$.
	\item With input $\nu \in R^{n-\rho}$, the system
\begin{equation}\label{closed-2}
  \dot{\vartheta}=f(\vartheta,\gamma(\eta + \nu,\xi))
\end{equation}
is input-to-state stable (ISS), which implies that the origin of the closed-loop system \eqref{closed} is globally asymptotically stable.
\end{enumerate}
\end{assumption}
\subsection{Extended Kalman Filter}
If $\xi$ and $\sigma = C_1(\xi,u) \eta$ where measured but $\eta$ was not measured,  we could have designed an extended Klaman filter to estimate $\eta$ based on the system
\begin{equation}\label{ekf1}
\dot{\eta}=A_1(\xi,u)\eta+\phi_0(\xi,u),\qquad
\sigma = C_1(\xi,u) \eta.
\end{equation}
The filter is given by
\begin{equation}\label{ekf2}
\dot{\hat{\eta}}=A_1(\xi,u)\hat{\eta}+\phi_0(\xi,u)+L_r(t)
  [\sigma-C_1(\xi,u)\hat{\eta}],
\end{equation}
where
\begin{equation}
L_r=P_r C_1^T R^{-1},\label{ekf3}
\end{equation}
$P_r$ is the symmetric solution of the Riccati equation
\begin{equation}
\dot{P}_r=A_1 P_r + P_r A_1^T+Q-P_r C_1^T R^{-1} C_1P_r ,\quad P_r(0)=P_0 > 0,\label{ekf4}
\end{equation}
$Q(t)$ and $P_0$ are symmetric matrices, and $Q(t)$ and $R(t)$ satisfy
\begin{equation}
0<q_1I_{n-\rho}\le Q(t)\le q_2I_{n-\rho}, \qquad  0<r_1\le R(t) \le r_2 \label{ekf5}
\end{equation}
for some positive constants $q_1$, $q_2$, $r_1$, and $r_2$. In this case, the control would be
\begin{equation}
u=\gamma(\hat{\eta},\xi).\label{ekf6}
\end{equation}
and the closed-loop system would be
\begin{align}
\dot\vartheta&=f(\vartheta,\gamma(\eta-\tilde{\eta},\xi)),\label{ekf7}\\
\dot{\tilde{\eta}}&=
[A_1(\xi,\gamma(\eta-\tilde{\eta},\xi))-L_r(t)C_1(\xi,
\gamma(\eta-\tilde{\eta},\xi))]\tilde{\eta},\label{ekf8}
\end{align}
where $\tilde{\eta}=\eta-\hat{\eta}$ is the estimation error.
\begin{assumption}\label{asp1}
The Riccati equation \eqref{ekf4} has a positive definite solution that satisfies
\begin{equation}
0<p_1I_{n-\rho}\le P_r^{-1}(t)\le p_2I_{n-\rho},\quad\forall\ t\ge 0. \label{ekf9}
\end{equation}
for some positive constants $p_1$ and $p_2$.
\end{assumption}

\newtheorem{remark}{Remark}
\begin{remark}\label{r2}
Examples of systems where this assumption is satisfied can be found in \cite{krener2003convergence}, \cite{praly2000output}, \cite{baras} and in the context of designing an adaptive high-gain observer in \cite{boizot2010adaptive}.
\end{remark}
\subsection{Saturation}
In the next section, we will utilize an extended high-gain observer to provide estimates of $\xi$ and $\sigma$ to be used in equations \eqref{ekf2}-\eqref{ekf4} and \eqref{ekf6}. Due to the peaking phenomenon of high-gain observers \cite{esfandiari1992output}, we need to saturate all the terms that depend on $\xi$ and $\sigma$. The saturation is done outside a compact set, which is chosen to include the trajectories of the system \eqref{ekf7}-\eqref{ekf8} for the desired compact set of initial conditions. Let $V_1(\vartheta)$ be smooth ISS Lyapunov function for  the system  \eqref{closed-2} that satisfies the inequalities
\begin{align}
   & \alpha_1 (\|\vartheta\|) \leq V_1(\vartheta) \leq \alpha_2 (\|\vartheta\|), \label{sat1}\\
   & \frac{\partial V_1}{\partial \vartheta} f(\vartheta,\gamma(\eta - \tilde{\eta},\xi)) \leq - \alpha_3(\|\vartheta\|), \ \forall \ \|\vartheta\| \geq \alpha_4(\|\tilde{\eta}\|), \label{sat2}
\end{align}
for all $\vartheta$ and $\tilde{\eta}$, with some class ${\cal K}_{\infty}$ functions $\alpha_1$ and $\alpha_2$ and class ${\cal K}$ functions $\alpha_3$ and $\alpha_4$. The existence of $V_1$ is guaranteed by the ISS converse Lyapunov theorem~\cite{sontag1995characterizations}.
 Let $V_2(t,\tilde{\eta}) = \tilde{\eta}^T P_r^{-1} \tilde{\eta}$. The derivative of $V_2$ along the trajectories of the system \eqref{ekf8} satisfies
 \begin{equation}\label{sat3}
   \dot{V}_2 = - \tilde{\eta}^T P_r^{-1} (Q + P_r C_1^T R^{-1} C_1 P_r) P_r^{-1} \tilde{\eta} \leq - k_1 V_2,
 \end{equation}
for some positive constant $k_1$. Define the compact sets $\Omega_1 = \{V_1(\vartheta) \leq c_1 \}$ and $\Omega_2 =  \{V_2(t,\tilde{\eta}) \leq c_2 \}$. With $c_1 \geq \alpha_2(\alpha_4(\sqrt{c_2/p_1}))$, the set $\Omega_1 \times \Omega_2$ is a positively invariant set of the closed-loop system \eqref{ekf7}-\eqref{ekf8}. This is so because, in view of \eqref{sat3}, $\dot{V}_2 < 0$ on the boundary $V_2 = c_2$ and, in view of \eqref{sat1} and \eqref{sat2}, $\dot{V}_1 < 0$ on the boundary $V_1 = c_1$. Let ${\cal L} \in R^n$ and ${\cal M}_0 \in R^{n-\rho}$ be any compact sets such that $\vartheta(0) \in {\cal L}$ and $\tilde{\eta}(0) \in  {\cal M}_0$. We can choose $c_1$ and $c_2$ large enough, subject to the constraint $c_1 \geq \alpha_2(\alpha_4(\sqrt{c_2/p_1}))$, such that ${\cal L}$ is in the interior of $\Omega_1$ and ${\cal M}_0$ is in the interior of $\Omega_2$. In this case, the solution of \eqref{ekf7}-\eqref{ekf8} belongs to $\Omega_1 \times \Omega_2$ for all $t \geq 0$. Let $M_{\xi}$ be a positive constant that satisfies
 $M_{\xi} > \max_{\Omega_1} \|\xi\|$ and $\psi$ be a smooth globally bounded function such that  $\psi(\xi) = \xi$ for $\|\xi\| \leq M_{\xi}$. Define $\hat{\gamma}(\eta,\xi)$ by $\hat{\gamma}(\eta,\xi) = \gamma(\eta,\psi(\xi))$. The function $\hat{\gamma}(\eta,\xi)$ is smooth, globally bounded in $\xi$, and $\hat{\gamma}(\eta,\xi) = \gamma(\eta,\xi)$ in $\Omega_1$. Similarly we define the functions $\hat{A}_1(\eta,\xi)$, $\hat{\phi}_0(\eta,\xi)$, $\hat{a}(\eta,\xi)$, and $\hat{C}_1(\eta,\xi)$ from the functions $A_1$, $\phi_0$, $a$, and $C_1$, respectively, by changing their $\xi$ entry to $\psi(\xi)$. Another function to be saturated is the function $\phi_1(\eta,\xi)$, which will appear later in the output feedback control. It is  the derivative of $C_1(\xi,u)\eta$ under the state feedback control $u=\gamma(\eta,\xi)$, i.e.,
\begin{align*}
\phi_1(\eta,\xi)&=\left[C_1(\xi,\gamma(\eta,\xi))  \begin{array}{c} \ \\ \ \end{array} \right.  \\
& \qquad +  \left. \eta^T \frac{\partial C_1^T(\xi,u)}{\partial u}(\xi,\gamma(\eta,\xi)) \frac{\partial \gamma}{\partial \eta}(\eta,\xi)  \right] \times \\ & \qquad [A_1(\xi,\gamma(\eta,\xi))\eta
+\phi_0(\xi,\gamma(\eta,\xi))]\\
&\;\quad+\frac{\partial[C_1(\xi,\gamma(\eta,\xi))\eta]}{\partial\xi}[A\xi+ B[C_1(\xi,\gamma(\eta,\xi))\eta\\
&\quad\quad+a(\xi,\gamma(\eta,\xi))]].
\end{align*}
Its saturated version, $\hat{\phi}_1$, is defined similarly. The variable $\sigma$ appears only as a driving input of the extended Kalman filter. It can be saturated by $M_{\sigma} \ {\rm sat}(\sigma/M_{\sigma})$ where ${\rm sat}(\cdot)$ is the standard saturation function and $M_{\sigma}$ satisfies $M_{\sigma} > \max_{\Omega_1} |C_1(\xi,\gamma(\eta,\xi)) \eta|$.

\section{Main Result}\label{s1}

\subsection{Output Feedback}\label{s34}
The output feedback control is taken as
\begin{equation}\label{satu}
  u = \hat{\gamma}(\hat{\eta},\hat{\xi})
\end{equation}
where $\hat{\eta}$ and $\hat{\xi}$ are provided by the observer
\begin{align}
\dot{\hat{\eta}} & = \hat{A}_1(\hat{\eta},\hat{\xi})\hat{\eta}+
 \hat{\phi}_0(\hat{\eta},\hat{\xi}) \nonumber\\
& \hspace*{0.5cm}  +L(t)\left[ M_{\sigma}\text{sat}(\hat\sigma/M_{\sigma})-
\hat{C}_1(\hat{\eta},\hat{\xi})\hat{\eta}\right],\label{ob233}\\
\dot{\hat{\xi}} & = A\hat{\xi}+ B[\hat{\sigma}+\hat{a}(\hat{\eta},\hat{\xi})]+H(\varepsilon)(y-C\hat{\xi}),\label{ob213} \\
\dot{\hat{\sigma}} & =\hat{\phi}_1(\hat{\eta},\hat{\xi})+(\alpha_{\rho+1}/\varepsilon^{\rho+1})
(y-C\hat{\xi}).\label{ob223}
\end{align}
The observer gain $H(\varepsilon)$ is given by $$H(\varepsilon)=[\alpha_1/\varepsilon,...,\alpha_\rho/\varepsilon^{\rho}]^T,$$ where the constants $\alpha_1,...,\alpha_{\rho +1}$ are chosen such that the polynomial
$ s^{\rho + 1}+\alpha_1 s^\rho + ... + \alpha_{\rho +1}$ is Hurwitz and $\varepsilon>0$ is a small parameter. The observer gain $L(t)$ is given by
\begin{equation}
L=P\hat{C}_1^TR^{-1}\label{l2}
\end{equation}
and $P(t)$ is the symmetric solution of the Riccati equation
\begin{equation}
\dot{P}=\hat{A}_1P+P\hat{A}_1^T+Q-P\hat{C}_1^TR^{-1}\hat{C}_1P,\quad P(0)=P_0> 0,\label{rec2}
\end{equation}
$P_0$ is a symmetric matrix and $Q(t)$ and $R(t)$ satisfy \eqref{ekf5}.
\begin{assumption}\label{asp}
The Riccati equation \eqref{rec2} has a positive definite solution that satisfies
\begin{equation}
0<p_3I_{n-\rho}\le P^{-1}(t)\le p_4I_{n-\rho}, \quad \forall t\ge 0,  \label{p43}
\end{equation}
for some positive constants $p_3$ and $p_4$.\footnote{Throughout the paper, the positive constants $a_i$, $b_i$, $c_i$, $k_i$, $L_i$, and $p_i$ are independent of $\varepsilon$.}
\end{assumption}
\begin{remark}\label{r2}
The right-hand side of the Riccati equation \eqref{rec2} coincides with the right-hand side of the Riccati equation \eqref{ekf4} except during the peaking period of the observer. Therefore, it is reasonable to expect \eqref{p43} to follow from \eqref{ekf9} for sufficiently small $\varepsilon$. Proving such relationship, however, would be quite involved. Therefore, we opted to state the property \eqref{p43} as a separate assumption.
\end{remark}
Let
\begin{align}
&\tilde{\eta}=\eta - \hat{\eta},\label{eta_er3}\\
&\chi_i=(\xi_i-\hat{\xi}_i)/ \varepsilon^{\rho+1-i},\quad \quad 1\le i\le \rho, \label{eq143}\\
&\chi_{\rho+1}=C_1(\xi,\gamma(\hat{\eta},\xi)) \eta-\hat{\sigma}, \label{eq153}
\end{align}
 $\varphi=[\chi_1,\chi_2,...,\chi_\rho]^T$, $ D(\varepsilon)=\text{diag}[\varepsilon^\rho,\varepsilon^{\rho-1},...,\varepsilon]$,
 $\chi=[\varphi^T\,\,\chi_{\rho+1}]^T$, and $D_1(\varepsilon)=\text{diag}[D,\, 1]$.
  Equation \eqref{eq143} can be written as
\begin{equation}
D(\varepsilon)\varphi=\xi-\hat{\xi},\label{xi-er}
\end{equation}
and
$$D_1(\varepsilon)\chi=\begin{bmatrix}
\xi-\hat{\xi}\\
C_1(\xi,\gamma(\hat{\eta},\xi))\eta-\hat{\sigma}
\end{bmatrix}.$$
Therefore, the closed-loop system under \eqref{satu} is given by
\begin{align}
\dot{\eta}=&A_1(\xi,\hat{\gamma}(\hat{\eta},\hat{\xi}))\eta+
\phi_0(\xi,\hat{\gamma}(\hat{\eta},\hat{\xi})),\label{slow13}\\
\dot{\xi}=&A\xi+B[C_1(\xi,\hat{\gamma}(\hat{\eta},\hat{\xi}))\eta+
a(\xi,\hat{\gamma}(\hat{\eta},\hat{\xi}))],\label{slow23}\\
\begin{split}
\dot{\tilde{\eta}}=&A_1(\xi,\hat{\gamma}(\hat{\eta},\hat{\xi})){\eta}
-\hat{A}_1(\hat{\eta},\hat{\xi})\hat\eta+
\phi_0(\xi,\hat{\gamma}(\hat{\eta},\hat{\xi}))\\
&-\hat{\phi}_0(\hat{\eta},\hat{\xi})-P\hat{C}_1^T(\hat{\eta},\hat{\xi})R^{-1}\times\\
&\;\,\left[M_{\sigma}\text{sat}\left(\frac{C_1(\xi,\gamma(\hat{\eta},\xi))
\eta-\chi_{\rho+1}}{M_{\sigma}}\right)-
\hat{C}_1(\hat\xi,\hat{\eta})\hat\eta\right],\label{slow33}
\end{split}\\
\varepsilon\dot{\chi}=&\Lambda\chi+\varepsilon[\bar{B}_1\Delta_1+
\bar{B}_2\Delta_2/\varepsilon],\label{fast3}
\end{align}
where $\Lambda$ is a Hurwitz matrix in an observable canonical form with $\alpha_1$ to $\alpha_{\rho+1}$ in the first column,  $\bar{B}_1= [0,B^T]^T$,
$\bar{B}_2=[B^T,0]^T$, and the locally Lipschitz functions $\Delta_1$ and $\Delta_2$ are given by
\begin{align*}
\Delta_1=&
\bar{\phi}_1(\eta,\xi,\hat{\eta},\hat{\xi},\varepsilon)-\hat{\phi}_1(\hat{\eta},\hat{\xi}),\\
\bar{\phi}_1=&
C_1(\xi,\gamma(\hat{\eta},\xi)) [A_1(\xi,\hat{\gamma}(\hat{\eta},\hat{\xi}))\eta+
\phi_0(\xi,\hat{\gamma}(\hat{\eta},\hat{\xi}))]\\
&   +   \eta^T \frac{\partial C_1^T(\xi,u)}{\partial u}(\xi,\hat{\gamma}(\hat{\eta},\hat{\xi})) \frac{\partial \gamma}{\partial \eta}(\hat{\eta},\hat{\xi})  \times \\ & \quad
[\hat{A}_1(\hat{\eta},\hat{\xi})\hat{\eta}+
 \hat{\phi}_0(\hat{\eta},\hat{\xi}) \\
& \quad  +L(t) [ M_{\sigma}\text{sat}(\hat\sigma/M_{\sigma})-
\hat{C}_1(\hat{\eta},\hat{\xi})\hat{\eta}] ]\\
& + \frac{[\partial C_1(\xi,\gamma(\hat{\eta},\xi))\eta]}{\partial\xi}\times\\
&\quad \left[A\xi+B[C_1(\xi,\hat{\gamma}(\hat{\eta},\hat{\xi}))\eta
+a(\xi,\hat{\gamma}(\hat{\eta},\hat{\xi}))]  \right], \\
\Delta_2=& a(\xi,\hat{\gamma}(\hat{\eta},\hat{\xi}))-\hat{a}(\hat{\eta},\hat{\xi})\\
&+C_1(\xi,\hat{\gamma}(\hat{\eta},\hat{\xi}))\eta-C_1(\xi,\gamma(\hat{\eta},\xi))\eta.
\end{align*}
It can be shown that for $\|\hat{\xi}\| \leq M_{\xi}$, $\|\hat{\sigma}\| \leq M_{\sigma}$,  and bounded $\xi$, $\eta$ and $\hat{\eta}$, the functions $\Delta_1$ and $\Delta_2/\varepsilon$ satisfy
\begin{equation}\label{bound}
  |\Delta_1| \leq k_2(\|\tilde{\eta}\| + \|\chi\|), \qquad |\Delta_2/\varepsilon| \leq k_3 \|\chi\|
\end{equation}
for some positive constants $k_2$ and $k_3$. The closed-loop system  \eqref{slow13}-\eqref{fast3} has an equilibrium point at the origin $(\eta=0,\xi=0,\tilde{\eta}= 0, \chi=0)$.

\subsection{Stability Recovery}\label{ss}
Define the sets of  initial states as $(\eta(0),\xi(0)) = \vartheta(0) \in {\cal L}$,  $\tilde{\eta}(0) \in {\cal M} = \{\|\tilde{\eta}\| \leq k_4\}$,  and $(\hat\xi(0),\hat{\sigma}(0))\in\mathcal{N}$, where   $\mathcal{N}$ is any compact subsets of $R^{\rho+1}$ and $k_4$  is constrained by $c_1 > \alpha_2(\alpha_4(k_4 \sqrt{p_4/p_3}))$. Recall that ${\cal L}$ is in the interior of $\Omega_1$ by the choice of $c_1$.
\newtheorem{theorem}{Theorem}
\begin{theorem}\label{t31}
Consider the closed loop system \eqref{slow13}-\eqref{fast3} under Assumptions \ref{ass1}-\ref{asp}. Then, there exists $\varepsilon_1^* > 0$, such that for all $0<\varepsilon<\varepsilon_1^*$, the origin of the closed loop system is asymptotically stable and $\mathcal{L}\times\mathcal{M}\times\mathcal{N}$ is a subset of its region of attraction. Moreover, if the origin of $\eqref{closed}$ is exponentially stable, so is the origin of \eqref{slow13}-\eqref{fast3}.
\end{theorem}

\begin{remark}\label{ra}
The sets ${\cal L}$ and ${\cal M}$ can be made arbitrarily large by choosing $c_1$ and $k_4$ large enough, subject to the constraint $c_1 > \alpha_2(\alpha_4(k_4 \sqrt{p_4/p_3}))$. Notice, however, that larger $c_1$ means larger saturation levels $M_{\xi}$ and $M_{\sigma}$.
\end{remark}

\subsection{Proof of Theorem 1}\label{a1}

For convenience, rewrite \eqref{slow13}-\eqref{slow23} as $\dot\vartheta\triangleq f_{a}(\vartheta,\tilde\eta,D\varphi)$ and \eqref{slow33} as $\dot{\tilde{\eta}}= f_{b}(\vartheta,\tilde{\eta},t,D_1\chi)$; note that $f_a(\vartheta,\tilde{\eta},0) = f(\vartheta,\gamma(\eta-\tilde{\eta},\xi))$.  Let $V_3(t, \tilde{\eta}) = \tilde{\eta}^T P^{-1} \tilde{\eta}$ and $W(\chi)=\chi^TP_0\chi$.
where $P_0$ is the positive definite solution of $P_0 \Lambda +\Lambda^T P_0=-I$.
Define the sets $\Omega_3 = \{V_3(t,\tilde{\eta}) \leq c_3 \}$ and $\Sigma=\{ W(\chi)\le\beta\varepsilon^2\}$, where $c_3$ satisfies $c_3 > p_4 k_4^2$ and $c_1 \geq \alpha_2(\alpha_4(\sqrt{c_3/p_3}))$,  and $\beta$ is a positive constant independent of $\varepsilon$ to be determined. Let  $\mathcal{S}=\Omega_1 \times \Omega_3 \times\Sigma$. In what follows, we prove that $\mathcal{S}$ is positively invariant and all trajectories starting in $\mathcal{L}\times\mathcal{M}\times\mathcal{N}$ enter $\mathcal{S}$ in finite time. Furthermore, by analysis inside this set we prove asymptotic stability of the origin. For this purpose, we notice that inside $\mathcal{S}$, $\vartheta \in \Omega_1$ and, for sufficiently small $\varepsilon$, $\|\hat{\xi}\| < M_{\xi}$ and $|\hat{\sigma}| < M_{\sigma}$. Therefore, $\psi(\hat{\xi}) = \hat{\xi}$,
$M_{\sigma} \ {\rm sat}(\hat{\sigma}/M_{\sigma}) = \hat{\sigma}$,   $\hat{\gamma}(\hat{\eta},\hat{\xi}) = \gamma(\hat{\eta},\hat{\xi})$, and $\hat{A}_1(\hat{\eta},\hat{\xi})= A_1(\hat{\xi},\gamma(\hat{\eta},\hat{\xi}))$. Similarly, $\hat{\phi}_0$, $\hat{a}$, $\hat{C}_1$, and $\hat{\phi}_1$ coincide with $\phi_0$, $a$, $C_1$, and $\phi_1$, respectively. 
Consequently, 
\[  f_b(\vartheta,\tilde{\eta},t,D_1 \chi) = [\hat{A}_1(\hat{\eta},\hat{\xi}) - L \hat{C}_1(\hat{\eta},\hat{\xi})]\tilde{\eta} + O(\|\chi\|). \]
For any $0<\varepsilon_1\le 1$, there are positive constants $L_1,$ to $L_5$, such that for all $(\vartheta, \tilde{\eta},\chi)\in\mathcal{S}$ and every $0<\varepsilon\le\varepsilon_1$ and $t\geq 0$, we have
\begin{align}
\vectornorm{f_{a}(\vartheta,\tilde\eta,D\varphi)-f_{a}(\vartheta,\tilde\eta,0)}&\le L_1\vectornorm{\chi},\label{fa}\\
\vectornorm{f_{b}(\vartheta,\tilde{\eta},t,D_1\chi)- f_{b}(\vartheta,\tilde{\eta},t,0)}&\le L_2\vectornorm{\chi},\label{fb}\\
\vectornorm{f_{a}(\vartheta,\tilde\eta,D\varphi)-f_{a}(\vartheta,0,D\varphi)}&\le L_3\vectornorm{\tilde{\eta}},\label{fa1}\\
|\Delta_1| \leq L_4, \qquad |\Delta_2| \leq L_5. \label{fa2}
\end{align}
The set $\mathcal{S}$ is positively invariant because for all $(\vartheta, \tilde{\eta}, \chi) \in \mathcal{S}$,
\begin{eqnarray*}
  \dot{V}_1 &=& \frac{\partial V_1}{\partial \vartheta} f_{a}(\vartheta,\tilde{\eta},0) + \frac{\partial V_1}{\partial \vartheta} [f_{a}(\vartheta,\tilde\eta,D\varphi) - f_{a}(\vartheta,\tilde{\eta},0)]\\
   & \leq & - \alpha_3(\|\vartheta\|) + k_5 \|\chi\|,
\end{eqnarray*}
on the boundary $V_1 = c_1$,
\begin{eqnarray*}
  \dot{V}_3 &=& \frac{\partial V_3}{\partial t} + \frac{\partial V_3}{\partial \tilde{\eta}}f_{b}(\vartheta,\tilde{\eta},t,0)\\
   && \mbox{} +\frac{\partial V_3}{\partial \tilde{\eta}}[f_{b}(\vartheta,\tilde{\eta},t,D_1\chi) -  f_{b}(\vartheta,\tilde{\eta},t,0)]  \\
   & \leq &  - k_6 V_3 + k_7 \|\chi\|,
\end{eqnarray*}
on the boundary $V_3 = c_3$, and
\begin{eqnarray*}
 \varepsilon \dot{W} &=& - \|\chi\|^2 + 2 \varepsilon \chi^T P_0 [\bar{B}_1\Delta_1+
\bar{B}_2\Delta_2/\varepsilon] \\
   & \leq & - k_8 W  + \varepsilon k_9 \sqrt{W}
\end{eqnarray*}
on the boundary $W = \beta \varepsilon^2$,
for some positive constants $k_5$ to $k_9$. Taking $\beta >  (k_9/k_8)^2$, we conclude that there is $\varepsilon_2 > 0$ such that for all $0 < \varepsilon \leq \varepsilon_2$,  $\dot{V}_1 < 0$ on $V_1 = c_1$, $\dot{V}_3 < 0$ on $V_3 = c_3$ and $\dot{W} < 0$ on  $W = \beta \varepsilon^2$. Thus, $\mathcal{S}$ is positively invariant for all $0<\varepsilon\le\varepsilon_2$.

Because ${\cal L}$ is in the interior of $\Omega_1$ and ${\cal M}$ in the interior of $\Omega_3$, we see that for all $\vartheta(0) \in \mathcal{L}$ and $\tilde{\eta}(0) \in \mathcal{M}$,  there is a finite time $T_0$ independent of $\varepsilon$ such that
$\vartheta(t) \in \Omega_1$ and $\tilde{\eta}(t) \in \Omega_3$, for all $ t\,\in[0,T_0]$.
During this time, the inequalities \eqref{bound} hold and $W$ satisfies
\begin{equation}\label{bound2}
  \varepsilon \dot{W} \leq - (1 - k_a \varepsilon) k_8 + \varepsilon k_b \sqrt{W}
\end{equation}
for some positive constants $k_a$ and $k_b$.
Even though $\chi(0)$ could be of the order of $1/\varepsilon^{\rho}$, \eqref{bound2} shows that   there exist $\varepsilon_3>0$ and $T(\varepsilon)>0$ with $T(\varepsilon)\to 0\,\, \text{as}\,\,\varepsilon\to 0$, such that $W(\chi(T(\varepsilon)))\le\beta\varepsilon^2$ for every $0<\varepsilon<\varepsilon_3$. Therefore, taking $\varepsilon_0^*=\text{min}\{\varepsilon_1,\varepsilon_2,\varepsilon_3\}$ ensures that the trajectory enters the set $\mathcal{S}$ during the time interval $[0,T(\varepsilon)]$, for every $0<\varepsilon<\varepsilon_0^*$, and does not leave thereafter.

We now work inside the set $\mathcal{S}$ to prove asymptotic stability of the origin. To this end, we recall that the origin of
$\dot{\vartheta}=f(\vartheta,\gamma(\eta,\xi))$ is globally asymptotically stable.
By the converse Lyapunov theorem [\cite{khalil}, Th. 4.17], there are a smooth positive definite radially unbounded function $V_4(\vartheta)$, and a class ${\cal K}$ function $\alpha$, such that
\begin{equation}
\frac{\partial V_4}{\partial \vartheta}f_a(\vartheta,0,0)\le-\alpha(\|\vartheta\|),\quad\forall\,\vartheta\in R^n\label{v13}
\end{equation}
Consider the Lyapunov function candidate $
V_5(t,\vartheta,\tilde\eta,\chi)=V_4(\vartheta)+\theta\sqrt{\tilde{\eta}^T P^{-1}\tilde{\eta}}+\sqrt{W(\chi)}$,
where $\theta>0$ is to be determined. Using the continuous differentiability of $V_4$, the Lipschitz properties of $f_a$, $f_b$, and \eqref{bound}, it can be shown that, for all $(\vartheta, \tilde{\eta},\chi)\in \mathcal{S}$,
\begin{align}
\begin{split}
\dot{V}_5\le&-\alpha(\vectornorm{\vartheta})-[a_1\theta-a_2]\vectornorm{\tilde{\eta}}\\
&-[a_3(\frac{1}{\varepsilon}-a_4)-\theta a_5-a_6]\vectornorm{\chi},
\end{split}
\end{align}
where $a_1$ to $a_6$ are positive constants. Choosing $\theta>a_2/a_1$ and $0<\varepsilon^*_1\le\varepsilon^*_0$ such that $\varepsilon^*_1<1/((\theta a_5+a_6)/a_3+a_4)$, then, for all $0<\varepsilon\le\varepsilon^*_1$, $\dot{V}_5$ is negative definite, which proves asymptotic stability.

Consider now the case when the origin of $\eqref{closed}$ is exponentially stable.
By the converse Lyapunov theorem~[\cite{khalil}, Th. 4.14], There exist a ball $B(0,r_1)$, for some $r_1>0$, and  a Lyapunov function $V_6$ that satisfies the following inequalities for all $\vartheta\in B(0,r_1)$
\begin{align}
&b_1\vectornorm{\vartheta}^2\le V_6(\vartheta)\le b_2\vectornorm{\vartheta}^2,\label{b23}\\
&\frac{\partial V_7}{\partial\vartheta}f(\vartheta,\gamma(\eta,\xi))\le-b_3\vectornorm{\vartheta}^2,\label{b33}\\
&\vectornorm{\frac{\partial V_6}{\partial\vartheta}}\le b_4\vectornorm{\vartheta},\label{b43}
\end{align}
for some positive constants $b_1, b_2, b_3$ and $b_4$. Consider the composite Lyapunov function $V_7(t,\vartheta,\tilde{\eta})=V_6(\vartheta)+\theta_1\tilde{\eta}^TP^{-1}\tilde{\eta}+W(\chi)$ with $\theta_1>0$. Choose $r_2< r_1$, then it can be shown, using \eqref{bound}, \eqref{b33}, \eqref{b43}, and Lipschitz properties of $f_a$ and $f_b$, that, for all $(\vartheta, \tilde{\eta}, \chi)\in B(0,r_2)\times \{\vectornorm{\tilde{\eta}}\le r_2\}\times\{\vectornorm{\chi}\le r_2\}$,
\begin{align*}
\dot{V_7}\le&-b_3\vectornorm{\vartheta}^2+c_4\vectornorm{\vartheta}\vectornorm{\tilde{\eta}}+c_5\vectornorm{\vartheta}\vectornorm{\chi}-\theta_1c_6\vectornorm{\tilde{\eta}}^2\\
&+\theta_1c_7\vectornorm{\tilde{\eta}}\vectornorm{\chi}-[\frac{1}{\varepsilon}-c_8]\vectornorm{\chi}^2+c_9\vectornorm{\tilde{\eta}}\vectornorm{\chi}\\
\le&-\begin{bmatrix}
\vectornorm{\vartheta}\\
\vectornorm{\tilde\eta}\\
\vectornorm{\chi}\end{bmatrix}^T\begin{bmatrix}
b_3&-c_4/2&-c_5/2\\
-c_4/2&\theta_1c_6&-(\theta_1c_7+c_9)/2\\
-c_5/2&-(\theta_1c_7+c_9)/2&[1/\varepsilon-c_8]
\end{bmatrix}\\
&\quad\times\begin{bmatrix}
\vectornorm{\vartheta}\\
\vectornorm{\tilde\eta}\\
\vectornorm{\chi}\end{bmatrix}
\triangleq-\begin{bmatrix}
\vectornorm{\vartheta}\\
\vectornorm{\tilde\eta}\\
\vectornorm{\chi}\end{bmatrix}^T\Gamma\begin{bmatrix}
\vectornorm{\vartheta}\\
\vectornorm{\tilde\eta}\\
\vectornorm{\chi}\end{bmatrix},
\end{align*}
where $c_4$ to $c_9$ are positive constants. Choose $\theta_1$ to make $\theta_1b_3c_6-c_4^2/4$ positive and $\varepsilon_2^*>0$ small enough such that, for all $0<\varepsilon\le\varepsilon_2^*$, the determinant of $\Gamma$ is positive. This makes $\Gamma$ positive definite, which completes the proof.  $\hfill\Box$

\subsection{Trajectory Recovery}
\begin{theorem}\label{T2}
Suppose the conditions of Theorem~\ref{t31} hold and let $(\vartheta_r(t),\tilde{\eta}_r(t))$ denote the the solution of \eqref{ekf7}-\eqref{ekf8} that starts from the same initial conditions  $(\vartheta(0),\tilde{\eta}(0))$ as the closed-loop system \eqref{slow13}-\eqref{fast3}. Given any $\mu>0$, there exists $\varepsilon_3^*>0$ such that, for every $0<\varepsilon\le\varepsilon_3^*$, we have
\begin{equation}\label{recov}
\vectornorm{\vartheta(t,\varepsilon)-\vartheta_r(t)}\le\mu, \quad
\vectornorm{\tilde{\eta}(t,\varepsilon)-\tilde{\eta}_r(t)}\le\mu
\end{equation}
for all $t\geq 0$.
\end{theorem}
\subsection{Proof of Theorem 2}\label{a2}
We start by noting that the origin $(\vartheta,\tilde{\eta})=(0,0)$ of \eqref{ekf7}-\eqref{ekf8} is globally uniformly  asymptotically stable. This is the case because the
system \eqref{ekf7} is input-to-state stable and the origin of \eqref{ekf8} is globally exponentially stable; see~[\cite{khalil}, Lemma 4.7]. Since both $(\vartheta(t,\varepsilon),\tilde{\eta}(t,\varepsilon))$ and $(\vartheta_r(t),\tilde{\eta}_r(t))$ converge to zero as $t$ tends to infinity, uniformly in $\varepsilon$, there is time $T_2 > 0$ such that \eqref{recov} holds for all $t \geq T_2$. So we are left with the task of proving \eqref{recov} on the compact interval $[0,T_2]$.

The closed loop system \eqref{slow13}-\eqref{fast3} with the Riccati equation \eqref{rec2} is in the singularly perturbed form, where \eqref{slow13}-\eqref{slow33} and \eqref{rec2} constitute the slow dynamics and \eqref{fast3} constitutes the fast dynamics. Let $\zeta = (\vartheta, \tilde{\eta}, P)$ denote the slow variables. We observe that the system formed of \eqref{ekf7}-\eqref{ekf8} and the Riccati equation~\eqref{ekf4}, with state $\zeta_r = (\vartheta_r, \tilde{\eta}_r, P_r)$, is nothing but the reduced system of \eqref{slow13}-\eqref{fast3} and \eqref{rec2} when $\varepsilon = 0$; this is so because $\varepsilon = 0$ yields $\chi=0$, which gives $\varphi=0$.
We know from the proof of Theorem \ref{t31} that, there exists $\varepsilon_0^*>0$ such that, for every $0<\varepsilon\le\varepsilon_0^*$, the trajectories are inside the set $\mathcal{S}$ for all $t\ge T(\varepsilon)$, where $\mathcal{S}$ is $O(\varepsilon)$ in the direction of the variable $\chi$. We divide the interval $[0,T_2]$ into two intervals $[0,T(\varepsilon)]$ and $[T(\varepsilon),T_2]$ and show \eqref{recov} for each interval.

1- The interval $[0,T(\varepsilon)]$\\
Since $P(t)$ is bounded and all right-hand side functions are globally bounded in $\hat{\xi}$ and $\hat{\sigma}$, there is a a constant $b_5$ such that
\[ \|\zeta(t,\varepsilon) - \zeta(0)\| \leq b_5 t  \]
Similarly, there is a constant $b_6$ such that
\[ \|\zeta_r(t) - \zeta_r(0)\| \leq b_6 t  \]
Since $\zeta(0) = \zeta_r(0)$,
\[ \|\zeta(t,\varepsilon) - \zeta_r(t)\| \leq 2 b_7 T(\varepsilon) \]
for all $t \in [0, T(\varepsilon)]$, where $b_7 = \max\{b_5, b_6\}$.
Since $T(\varepsilon)\to 0$ as $\varepsilon\to 0$, there exists $0<\varepsilon_4\le\varepsilon_1^*$ such that \eqref{recov} holds on $[0, T(\varepsilon)]$ for every $0<\varepsilon\le\varepsilon_4$.

2- The interval $[T(\varepsilon),T_2]$ \\
Inequality~\eqref{recov} follows from continuous dependence of the solution of differential equations on initial states and right-hand side functions~[\cite{khalil}, Th. 3.5]
since the right-hand side of the singularly perturbed system is $O(\varepsilon)$ close to the right-hand side of the reduced system and $\|\zeta(T(\varepsilon),\varepsilon) - \zeta_r(T(\varepsilon))\| \leq 2 b_7 T(\varepsilon)$. $\hfill\Box$

\begin{remark}
Theorem \ref{T2} shows that the output feedback control that uses the extended high-gain observer \eqref{ob213}-\eqref{ob223} can asymptotically recover the trajectories of the system \eqref{ekf7}-\eqref{ekf8}. This is beneficial as, first, it allows us to assume that the states $\xi,\eta$ and $\tilde\eta$ are available in the state-feedback design stage, and thus, simplifies the design procedure. Moreover, it  allows for tuning the control $u$ to satisfy certain performance requirements in the state-feedback-design stage. This can be done by simulating equations \eqref{ekf7}-\eqref{ekf8} and checking if the trajectories satisfy the design specifications. Second, it shows that the control design has to take into consideration the estimation error $\tilde\eta$, which is considered as an input to the closed-loop state feedback system. Therefore, by using equation $\eqref{ekf8}$, one can deduce the properties of this input under output feedback, and hence, can design a state feedback control accordingly.
\end{remark}
\section{Simulation Example}\label{ex}
Consider the single-input, single-output system
\begin{align}
\dot \eta=\xi+\eta\cos \xi	,\quad \dot \xi=\xi^2+\eta+u,\quad
y=\xi,
\end{align}
which is a special case of \eqref{13}-\eqref{33} with $A_1 = \cos \xi$, $\phi_0 = \xi$, $C_1 = 1$, and $a = \xi^2 + u$.
The system is non-minimum phase because the zero dynamics are $\dot \eta=\eta$.

We  design a globally stabilizing state feedback control using backstepping \cite{khalil}. Accordingly, it can be shown that the control
\begin{equation}
u=-4\eta-3\xi-\xi^2-2\eta\cos\xi
\end{equation} 
satisfies Assumption \ref{ass23} and renders the equilibrium point $(\eta_,\xi)=(0,0)$ globally exponentially stable.

The full order observer is designed according to the procedure in Section \ref{s34} but with a modification since $y = \xi$. We take the functions $A_1$, $\phi_0$ and $a$ in terms of y instead of $\hat{\xi}$. Therefore, we do not need to saturate $\hat{\xi}$. The observer and control are given by
\begin{align}
&\dot{\hat{\xi}}= y^2+\hat\sigma+u+\frac{\alpha_1}{\varepsilon}[y-\hat \xi],\\
&\dot{\hat{\sigma}}=\phi_1(\hat \eta,y)+\frac{\alpha_2}{\varepsilon^2}[y-\hat \xi],\\
&\dot{\hat{\eta}}= y+\hat \eta\cos y +L(t)[\hat\sigma-\hat \eta],\\
& u=-4\hat{\eta}-3 y- y^2-2 \hat{\eta} \cos y,
\end{align}
where $\phi_1(\hat \eta,y)=y+\hat \eta\cos y$. The observer gain $L$ is given by
$L = 0.1 P$, where $P$ is the solution of the Riccati equation
\begin{equation}
\dot{P} = 2 P \cos y + 1 -0.1 P^2, \quad P(0) = 0.1, \label{p}
\end{equation}
which corresponds to the choice $Q=1$ and $R=10$. The other observer parameters and initial conditions are taken as $\alpha_1=5$, $\alpha_2=1$, $\varepsilon=0.001$, $\eta(0)=0.5$, $\xi(0)=0.9$, $\hat \eta(0)=0$, $\hat \xi(0)=0.1$, and $\hat \sigma(0)=0$. Suppose that, under state feedback, $\eta$ belongs to the set $[-7,7]$. To guard against peaking, we  saturate $\hat{\sigma}$ at $\pm 10$.

\begin{figure}[t]
        \centering
        \includegraphics[width=\columnwidth, height=6 cm]{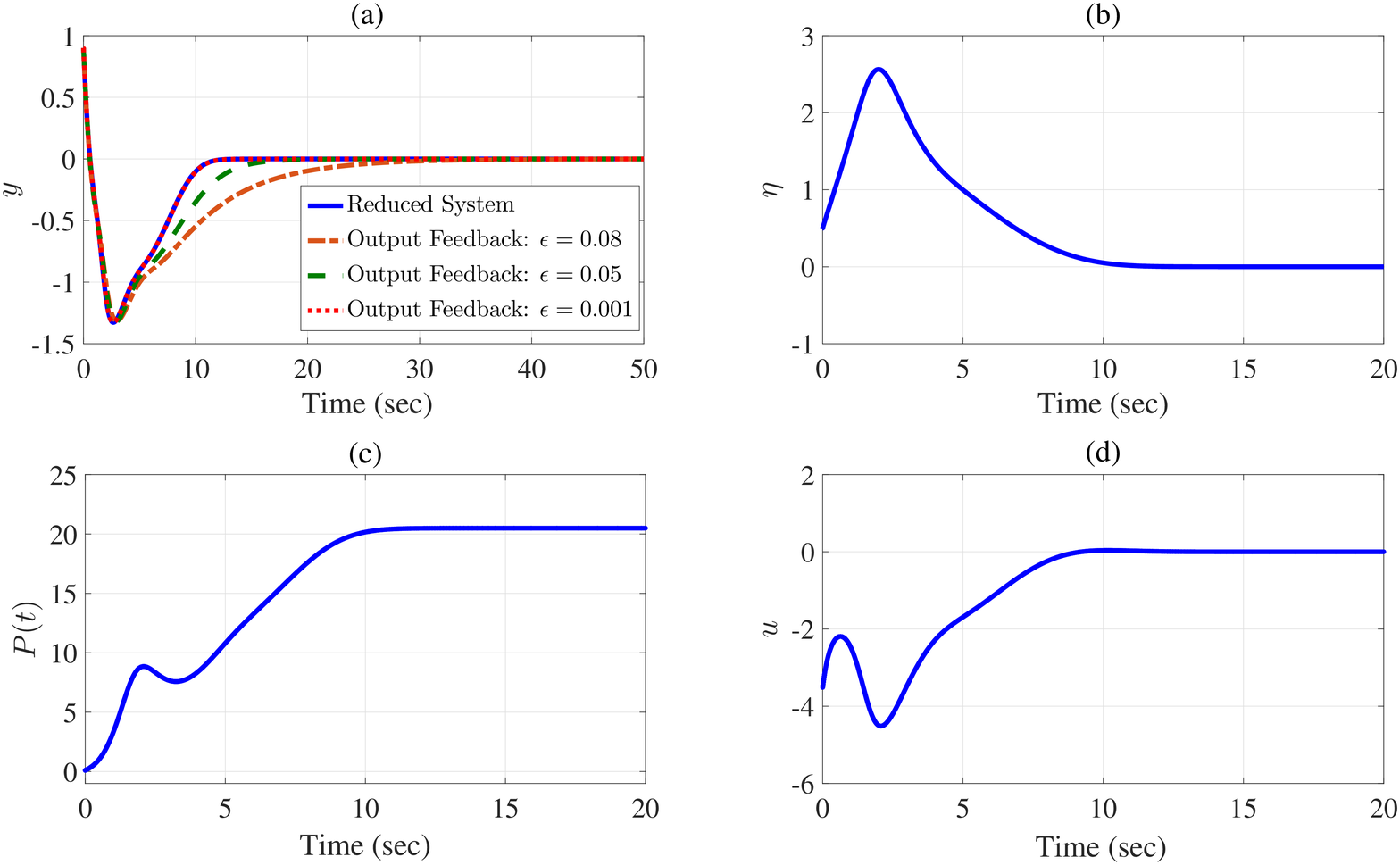}
        \vspace{-0.8 cm}
        \caption{(a) Comparison between the trajectories of $y$ of the reduced system and of the output feedback system for different values of $\varepsilon$. (b) Response of the state $\eta$ of the system under output feedback. (c) The trajectory of the Riccati solution $P(t)$. (d) Control effort.}
        \label{f1}
     \end{figure}

The trajectories of the output $y$ and the state $\eta$ of the system under output feedback are shown in Fig. \ref{f1}(a) and \ref{f1}(b). Fig. \ref{f1}(c) shows the trajectory of \eqref{p} over time. It clearly shows that $P(t)$ converges to a constant value, indicating the satisfaction of Assumptions \ref{asp1} and \ref{asp}. Fig. \ref{f1}(d) shows the control effort. Also shown in Fig. \ref{f1}(a) a comparison of the output $y = \xi$ of the system under output feedback  with $\xi$ of the reduced system \eqref{ekf7}-\eqref{ekf8} for different values of $\varepsilon$ and starting from the same initial conditions. It is clear from the figure that the smaller we choose $\varepsilon$ the closer the trajectories under output feedback become to those of the reduced system.

\section{Conclusions}\label{s6}
We presented an output feedback control strategy for nonlinear systems that could be non-minimum phase. The proposed controller is based on the nonlinear observer proposed in \cite{boker2013nonlinear}. The strategy allows for the use of any globally asymptotically stabilizing state feedback controller that satisfies an ISS condition when the estimation error of the internal dynamics is viewed as an input. We proved semi-global stabilization of the origin of the closed-loop system of the considered class of systems. We also showed that the output feedback control system can recover the performance of an auxiliary system that is composed of the system under partial state feedback with an extended Kalman filter that provides estimates of the unavailable states. In addition to this feature, which can prove useful in shaping the transient response of the system, the design procedure is relatively simple. Finally, we provided an example that demonstrates the effectiveness of the proposed control system.




%

\bibliographystyle{IEEEtran}
\bibliography{ref_acc_journal}
%
%

%





\end{document}